\documentclass[%
 reprint,
 amsmath,amssymb,
 aps,
]{revtex4-2}

\usepackage{graphicx}
\usepackage{dcolumn}
\usepackage{bm}

\begin{document}

\preprint{APS/123-QED}

\title{Field enhancement of intense laser pulses in a subwavelength plasma aperture}

\author{Xiaohui Gao}
  \email{gaoxh@utexas.edu}
\affiliation{Department of Physics, Shaoxing University, Shaoxing, Zhejiang 312000, China}

\date{\today}

\begin{abstract}
The interaction of intense, ultra-short laser pulses with nanostructures offers promising avenues for spatiotemporal light control. While enhanced optical transmission through subwavelength apertures has been extensively studied in the linear regime, its extension to ultrashort, high-intensity pulses remains largely unexplored. Here we demonstrate, through three-dimensional particle-in-cell simulations, significant field enhancement of intense laser pulses in subwavelength plasma apertures. The enhancement exhibits a non-resonant character, remaining robust across a wide range of plasma densities and saturating above approximately $20n_c$, while showing minimal dependence on wall thickness. Analysis of the Poynting vector reveals that energy concentration arises from interference between the incident field and back-scattered longitudinal field components. This size-dependent enhanced transmission in plasma apertures enables potential applications such as plasma-based dichroic filters operating at extreme intensities.
\end{abstract}

\maketitle
\section{Introduction}
The interaction of intense laser pulses with nanostructures constitutes an emerging frontier in laser physics~\cite{Ostrikov2016RMP, Rocca2024O, DombiRMP2020}, enabling a range of compelling phenomena including high-harmonic and X-ray generation~\cite{Rajeev2003PRL}, sub-cycle electron acceleration~\cite{Herink2012N}, controlled ion emission~\cite{Sun2024NC, Shou2025NC}, and advanced plasma optics~\cite{Peng2019,Xu2024MaRaE}. Subwavelength apertures, or nanoholes, serve as fundamental building blocks of metallic nanostructures, with their linear optical properties—particularly extraordinary optical transmission through a single hole or periodic arrays—being extensively studied in plasmonics~\cite{Genet2007N, GarciaVidal2010RoMP}. The significant field enhancement and confinement observed in single nanoholes are typically attributed to one of several interrelated mechanisms: localized Fabry-Perot-like resonances within the hole~\cite{CarreteroPRB}, impedance-matching conditions where the aperture acts as a resonant cavity~\cite{MedinaAPL2009}, or excitation of localized surface plasmons at the hole's rim~\cite{Degiron2004OC, GarciaVidal2010RoMP}.

The interaction of nanoholes with intense, ultra-short laser pulses differs from these conventional regimes in several aspects. In contrast to the linear response of noble metals at moderate intensities, intense laser irradiation induces a transition to a plasma state, fundamentally altering the electron density distribution and conductivity of the material. In addition, as the intensity reaches relativistic intensities, plasma dynamics and light-matter interactions become highly nonlinear. Furthermore, the interaction with few-cycle laser pulses introduces additional complexity, as the transient response and broad spectral bandwidth may significantly modify resonant-like energy coupling and field dynamics.

In this work, we employ particle-in-cell simulations to demonstrate robust field enhancement in nanoscale plasma apertures irradiated by intense laser pulses. Our model system consists of carbon nanotubes with diameters of hundreds of nanometers, a scale that is experimentally feasible~\cite{Truong2010} and generalizable to other materials such as gold nanotubes~\cite{KohlPRB2011}. While the properties of modes in cylindrical metallic waveguides have been studied previously~\cite{Prade1994guided,Novotny1994PRE,Takahara1997OL}, we present a comprehensive analysis of the complete physical process: field excitation at the aperture entrance, energy coupling into guided modes, and propagation through the plasma channel.

\section{Simulation Setting}
Three-dimensional particle-in-cell (PIC) simulations are performed using the open-source code \textsc{Smilei}~\cite{Derouillat2018CPC}. The simulation domain spans $8\lambda \times 3\lambda \times 3\lambda$, where $\lambda = 800$\,nm is the central laser wavelength. Spatial discretization is set to $\lambda/80$ in all directions (10\,nm grid spacing), with temporal resolution of $T/145$, where $T$ is the laser period, corresponding to a Courant number of 0.956. We use a carbon density of $8.8\times10^{22}$\,cm$^{-3}$~\cite{Dmowski2012JPCC}, equivalent to $52n_c$ for $\lambda=800$\,nm, with $n_c$ being the critical density.

A carbon nanotube structure, aligned along the $x$-direction, is positioned at the simulation center $(x, y, z) = (0,0,0)$ with wall thickness of 100\,nm unless otherwise specified. The system is initialized with 64 unionized macro-particles per cell and no initial free electrons. A linearly polarized laser pulse (electric field along $y$) propagates along $x$-axis, with Gaussian temporal profile (FWHM of two optical cycles at intensity), zero carrier-envelope phase, and peak amplitude reaching $x=0$ at $t=0$. A uniform transverse field profile is implemented, justified by the sub-wavelength nanotube dimensions relative to typical laser spots.

Collisional ionization processes are excluded from the primary simulations. Preliminary investigations revealed that while collisional ionization increases the average ion charge state, the field enhancement factor saturates at electron densities achievable without this mechanism. Consequently, omitting collisional ionization provides equivalent physical insight while substantially reducing computational requirements.

Boundary conditions employ periodic boundaries in transverse directions, Silver-Müller absorbing boundaries longitudinally for fields, and particle removal at longitudinal boundaries. Radiation damping effects are neglected as their contribution is minimal within the investigated parameter range.

\section{Results and Discussion}
\begin{figure}[htbp]
\centering
\includegraphics[width=0.35\textwidth]{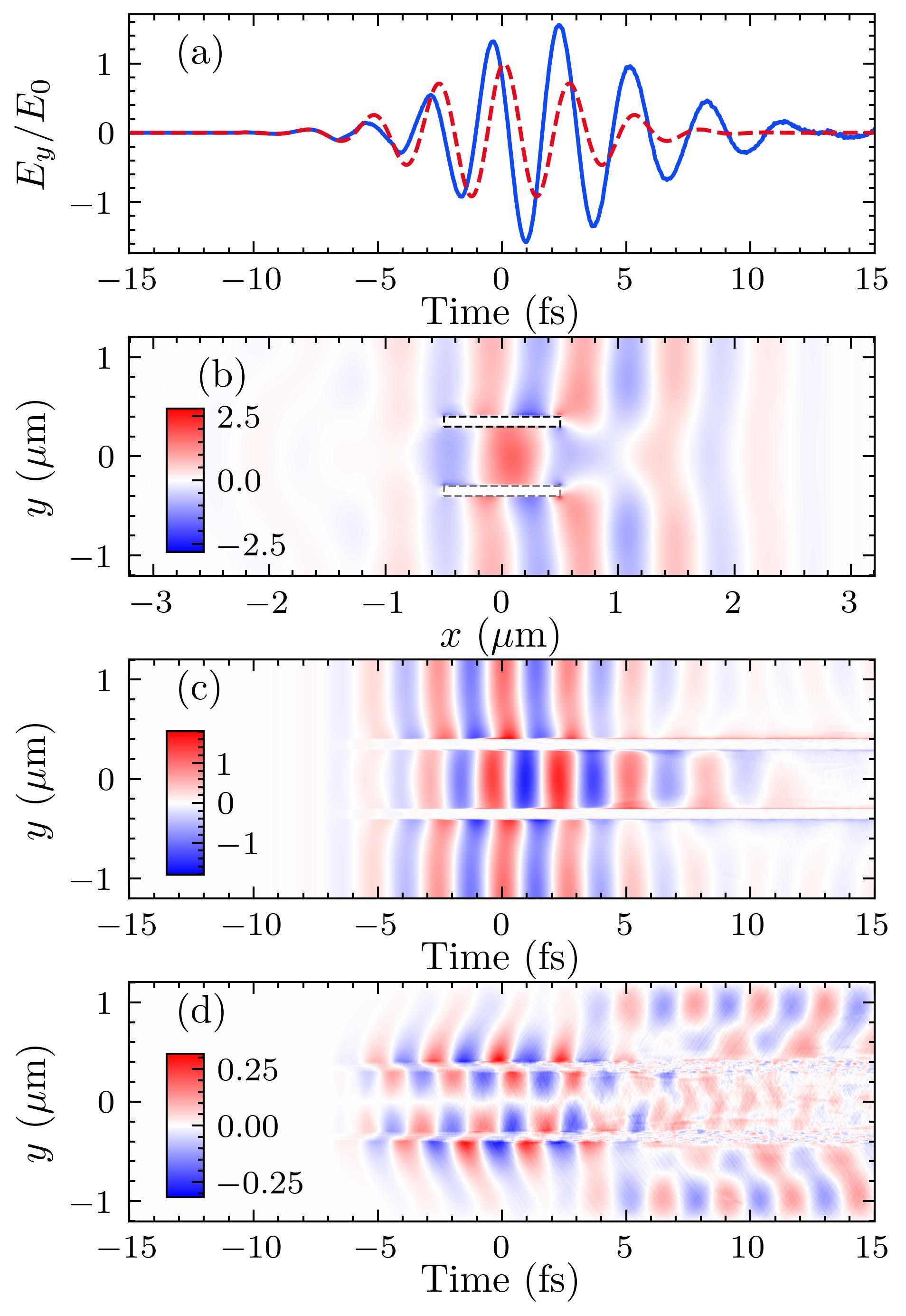}
\caption{Field enhancement in a carbon nanotube ($r_a = 300$\,nm) irradiated by a laser pulse ($I_0 = 10^{16}$\,W/cm$^2$). 
(a) Temporal profile of the transverse electric field $E_y$ at the nanotube center (solid blue) compared to the vacuum reference (dashed red), normalized to the incident field amplitude $E_0$. 
(b) Spatial distribution of $E_y$ in the $x$-$y$ plane at $t = 2.39$\,fs, with dashed lines indicating the nanotube boundary. 
(c) Time evolution of $E_y$ along the polarization axis $(x, z) = (0, 0)$. 
(d) Corresponding time evolution of the longitudinal field $E_x$ along the same axis.}
\label{fig1} 
\end{figure}%
Field enhancement in a carbon nanotube ($r_a = 300$\,nm) irradiated at $I_0 = 10^{16}$\,W/cm$^2$ is characterized in Fig.~\ref{fig1}. The normalized electric field $E_y/E_0$ at the nanotube center in Fig.~\ref{fig1}(a) demonstrates significant enhancement relative to vacuum reference, achieving peaks of approximately $\pm 1.5E_0$ during successive polarization cycles. The pulse duration within the nanotube nearly doubles, indicating substantial energy density increase through temporal compression. The snapshot of spatial field distribution in Fig.~\ref{fig1}(b) reveals remarkably uniform field profile throughout the nanotube core, suggesting efficient coupling into fundamental waveguide modes. Temporal evolution along the polarization axis in Figs.~\ref{fig1}(c) and~\ref{fig1}(d) reveals distinct field components: homogeneous transverse field $E_y$ throughout the core region and enhanced longitudinal components $E_x$ localized at top and bottom surfaces. The longitudinal field $E_x$ exhibits approximately one-third the amplitude of the transverse component and vanishes along the laser axis, while oscillating at higher frequency—behavior consistent with surface mode excitation.

\begin{figure}[htbp]
\centering
\includegraphics[width=0.4\textwidth]{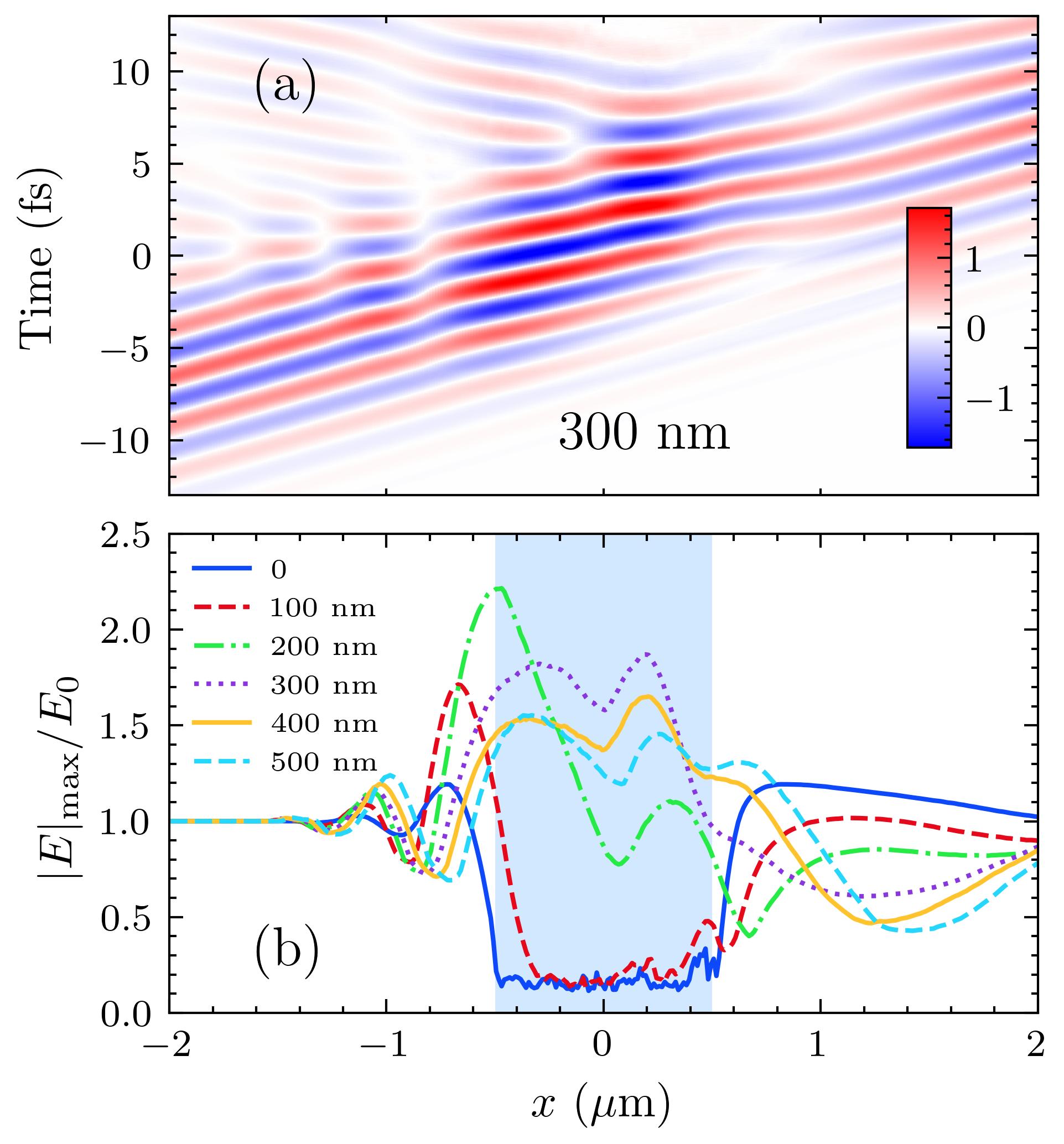}
\caption{Dependence of field enhancement on nanotube geometry. 
(a) On-axis temporal profile of $E_y$ versus axial position for $r_a = 300$\,nm. 
(b) Maximum normalized field amplitude along the laser axis for different inner radii, with light blue indicating positions inside the nanotube. The $r_a = 0$ case corresponds to a solid rod.}
\label{fig2} 
\end{figure}%
The geometric dependence of field enhancement reveals fundamental aspects of the underlying physical mechanism. As shown in Fig.~\ref{fig2}(a), the on-axis temporal field profile for $r_a = 300$\,nm shows characteristic wave reflection at both nanotube ends. Quantitative analysis across different radii in Fig.~\ref{fig2}(b) demonstrates that maximum enhancement factor increases with radius up to approximately $200$\,nm, beyond which it gradually decreases. The maximum field of $2.2E_0$ exceeds that of a standing wave formed by interference with reflection. 

This size dependence emerges from competing physical processes: efficient field enhancement at the nanotube entrance versus propagation losses within the plasma waveguide. For $r_a = 100$\,nm, the field decays rapidly to levels comparable to a solid rod within 0.2$\,\mu$m, consistent with cutoff conditions previously identified in metallic nanotubes~\cite{Novotny1994PRE}. In contrast, larger nanotubes ($r_a = 300$\,nm) exhibit significantly reduced attenuation, yielding optimal overall field enhancement through balanced entrance coupling and propagation characteristics.

\begin{figure}[htbp]
\centering
\includegraphics[width=0.4\textwidth]{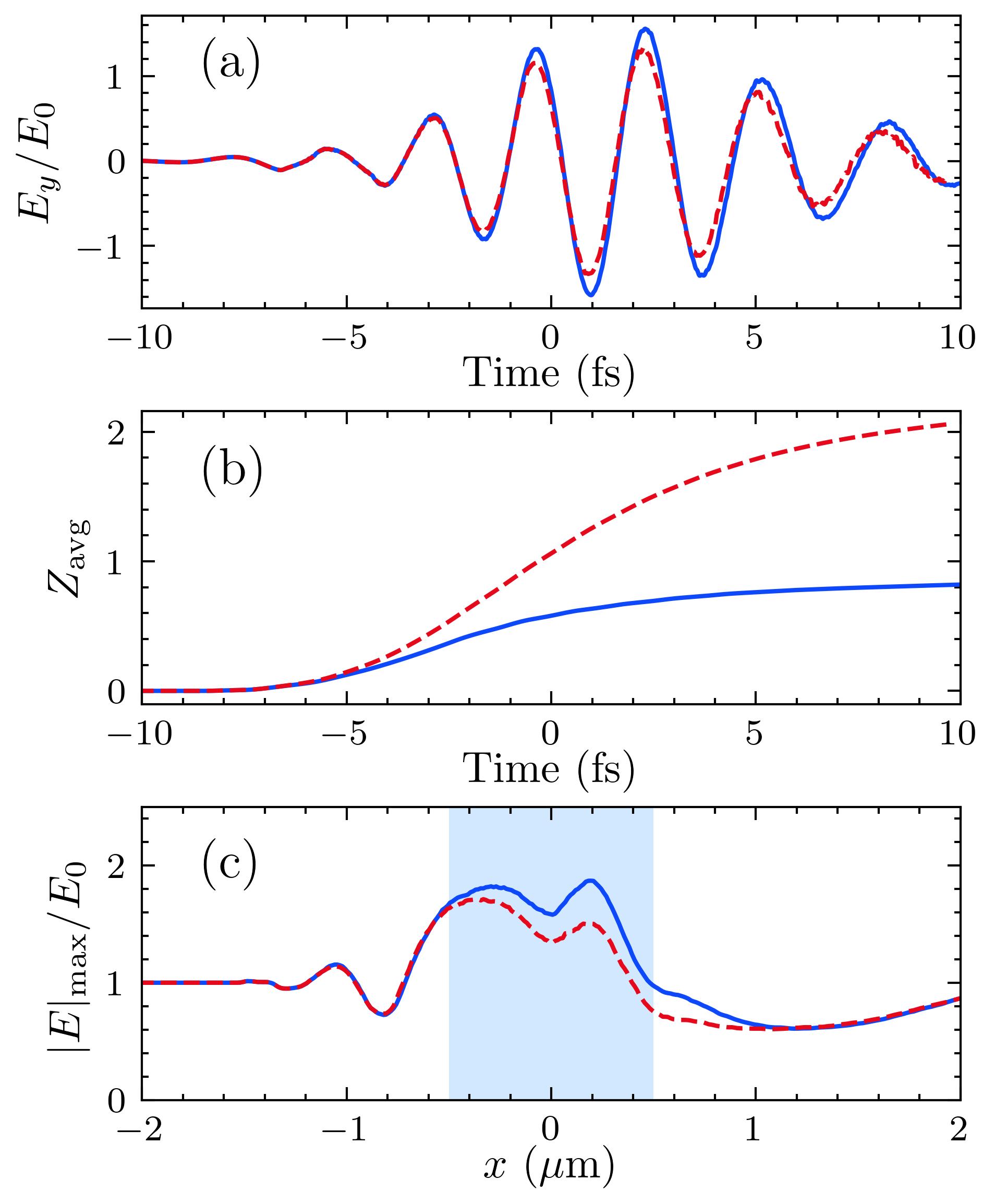}
\caption{Effect of collisional ionization on field enhancement in a carbon nanotube ($r_a = 300$\,nm). 
(a) Temporal evolution of the normalized electric field $E_y/E_0$ at the nanotube center. 
(b) Corresponding average charge state as a function of time. 
(c) Spatial profile of the maximum field amplitude along the longitudinal direction. 
Solid blue curves: simulations without collisional ionization; dashed red curves: with collisional ionization included.}
\label{figCI} 
\end{figure}%
The role of collisional ionization in field enhancement dynamics is examined in Fig.~\ref{figCI}. Comparison between baseline simulations and collisional ionization cases reveals that while collisional processes substantially increase the average charge state in Fig.~\ref{figCI}(b), they exert minimal influence on both the temporal field profile in Fig.~\ref{figCI}(a) and the peak field enhancement in Fig.~\ref{figCI}(c). Field distributions preceding the nanotube entrance remain nearly identical in both scenarios. However, significant discrepancies emerge during pulse propagation, with enhanced attenuation observed when collisional ionization is active. This increased propagation loss originates from higher electron density in the cladding plasma, intensifying waveguide dissipation through collisional damping.

\begin{figure}[htbp]
\centering
\includegraphics[width=0.45\textwidth]{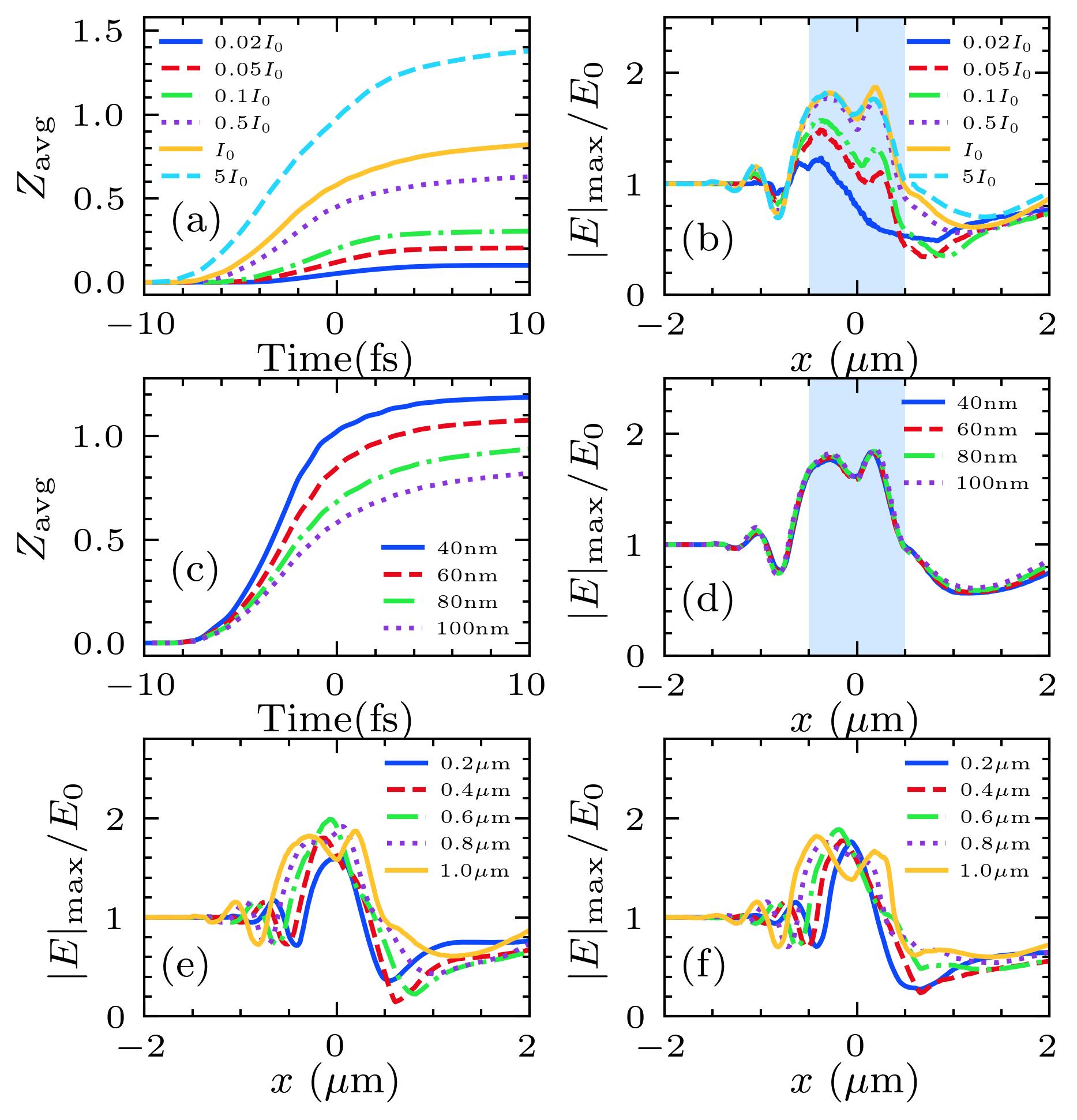}
\caption{Dependence of ionization dynamics and field enhancement on laser intensity and nanotube geometry. 
(a) Temporal evolution of average charge state for different laser intensities. 
(b) Spatial profile of maximum field amplitude for corresponding intensities. 
(c) Average charge state versus time for nanotubes with different wall thicknesses. 
(d) Maximum field amplitude versus position for different wall thicknesses.
(e) Spatial profile of the maximum normalized field amplitude $E_{\text{max}}/E_0$ along the nanotube axis for $I_0 = 10^{16}$ W/cm$^2$ (non-relativistic regime). Different curves correspond to nanotubes of varying length $L$. 
(f) Corresponding spatial profiles for $I_0 = 2\times10^{18}$ W/cm$^2$ (relativistic regime, $a_0 \approx 1$). The dashed vertical lines indicate the entrance ($x=0$) and exit positions of the nanotubes.}
\label{figPS} 
\end{figure}%
Comprehensive parameter studies in Fig.~\ref{figPS} reveal the robustness of field enhancement across varying operational conditions. The average charge state increases monotonically with laser intensity in Fig.~\ref{figPS}(a), while the field enhancement factor in Fig.~\ref{figPS}(b) exhibits initial growth followed by saturation above $0.5\times10^{16}$\,W/cm$^2$, corresponding to approximately 50\% ionization degree. Variation of wall thickness in Figs.~\ref{figPS}(c) and~\ref{figPS}(d) demonstrates that thicker walls reduce ionization through skin effect limitations, yet field enhancement remains largely unaffected—confirming that the fundamental mechanism is governed primarily by aperture geometry rather than wall properties. 

Analysis of nanotube length dependence in Figs.~\ref{figPS}(e) and~\ref{figPS}(f) reveals distinct behavior across intensity regimes. In the non-relativistic case, the shortest nanotube exhibits a single enhancement peak at the center, while longer structures develop double-peak distributions with maxima near both entrance and exit regions. At relativistic intensity ($a_0 \approx 1$), qualitative behavior persists with comparable enhancement factors, demonstrating mechanism robustness despite relativistic electron dynamics and nonlinear effects. The shell thickness is reduced to 40\,nm for computational efficiency at relativistic intensity. Despite of a higher ionization degree, the relativistic skin depth remains above the grid spacing due to the Lorentz factor. 

\begin{figure}[htbp]
\centering
\includegraphics[width=0.40\textwidth]{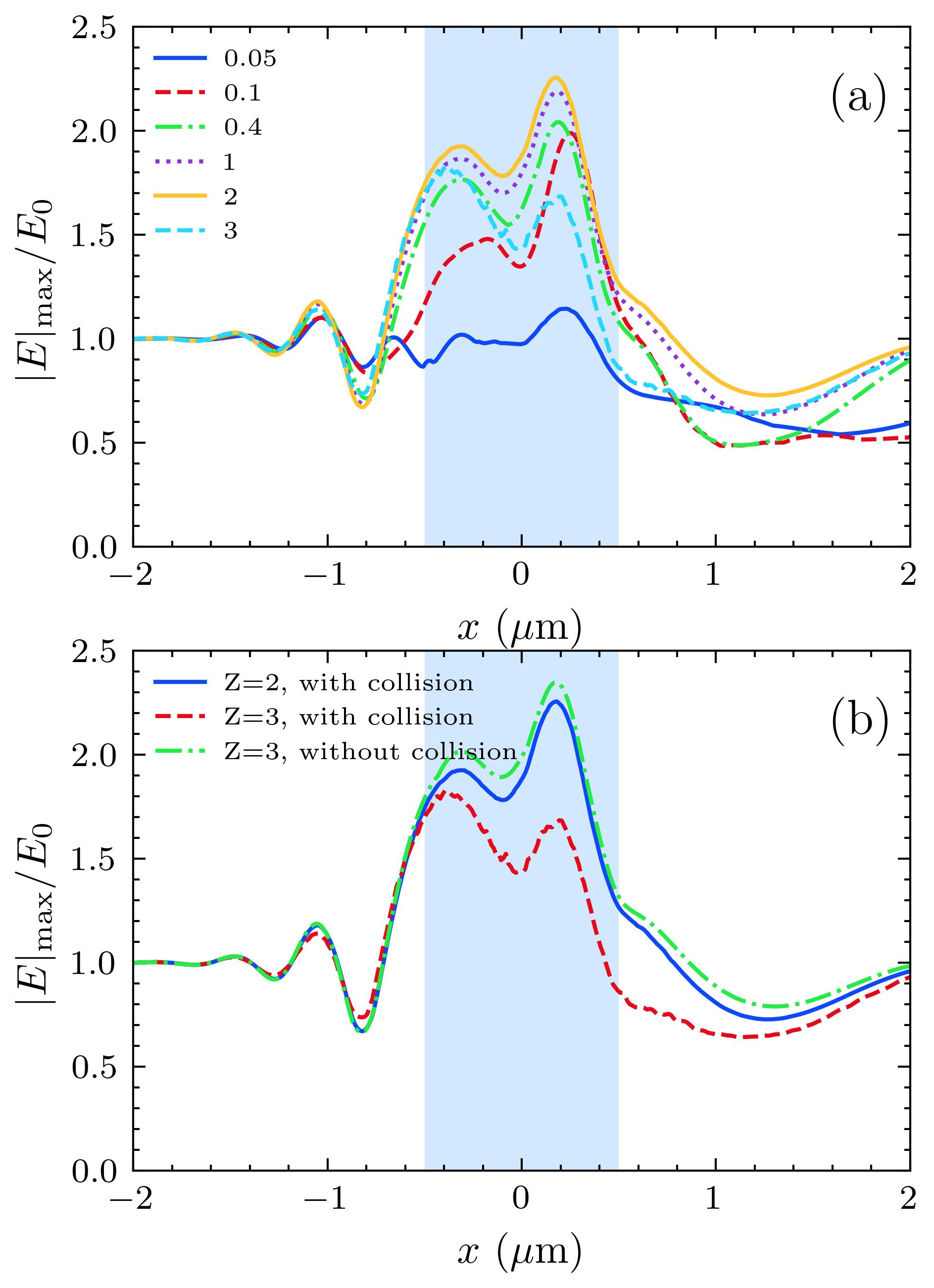}
\caption{Role of preionization and collisional effects in field enhancement. 
(a) Maximum normalized field amplitude along the longitudinal direction for tubular plasmas with different initial charge states. 
(b) Comparison of field enhancement with and without elastic electron-ion collisions.}
\label{figPRE} 
\end{figure}%
To distinguish between enhancement mechanisms originating from plasma aperture versus ionization dynamics, we conducted simulations with preionized plasmas in Fig.~\ref{figPRE}. Spatial field distributions for varying initial charge states in Fig.~\ref{figPRE}(a) maintain qualitative similarity to Fig.~\ref{fig2}, though enhancement is substantially greater in preionized cases due to eliminated ionization energy losses. The enhancement factor saturates at approximately charge state 0.4, corresponding to plasma density approximately 20$n_c$. Notably, enhancement decreases between $Z=2$ and $Z=3$. This is attributable to increased collisional damping. When elastic collisions are disabled in Fig.~\ref{figPRE}(b), the enhancement factor resumes increasing with $Z$, confirming collisional damping as the limiting mechanism at higher plasma densities.

\begin{figure}[htbp]
\centering
\includegraphics[width=0.4\textwidth]{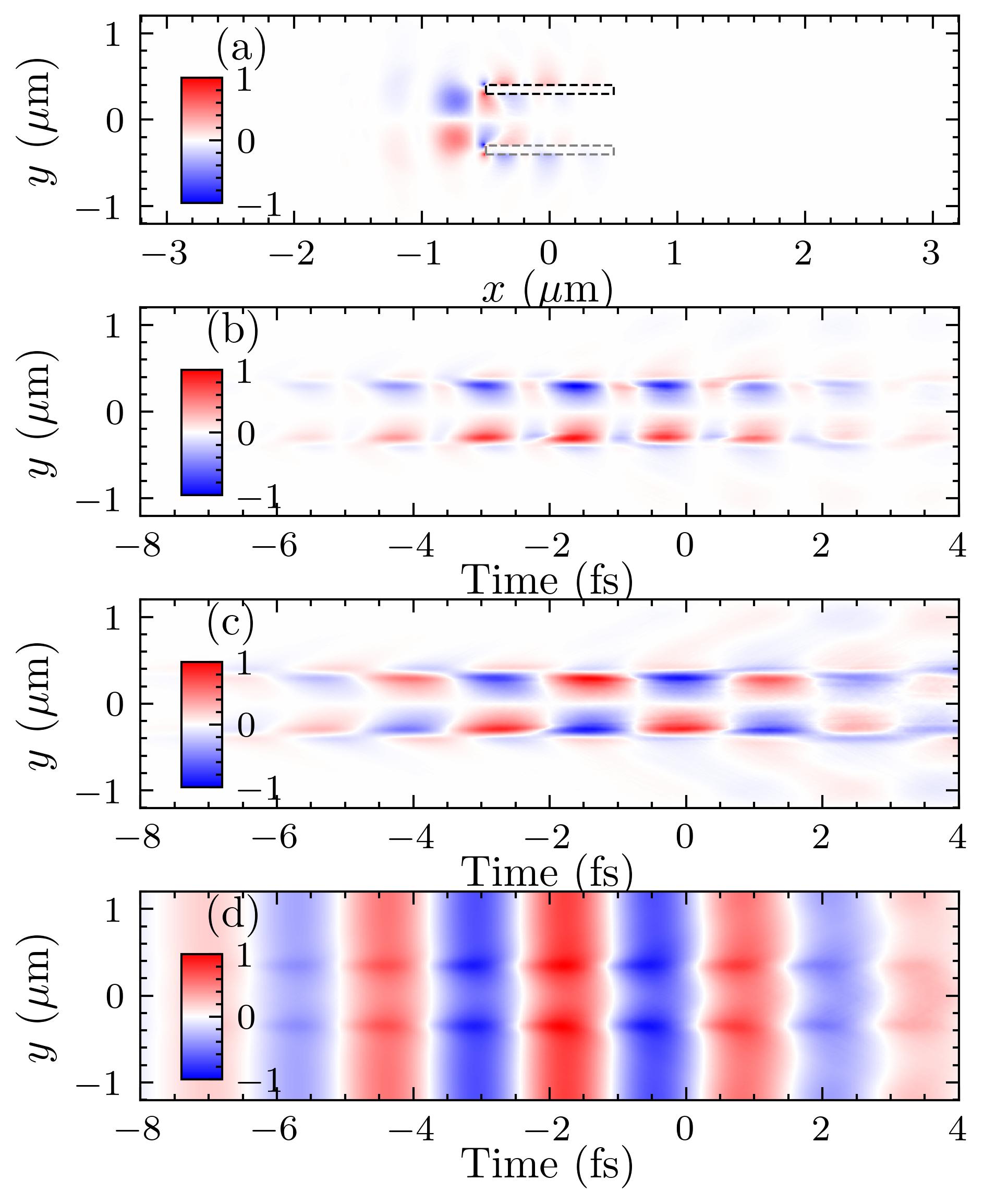}
\caption{Energy flow dynamics and field components near the nanotube entrance. 
(a) Spatial distribution of the $y$-component of the Poynting vector $P_y$ in the $x$-$y$ plane at $t=-3.86$ fs, showing energy convergence toward the central axis. 
(b) Time evolution of $P_y$ at position $(x,z)=(0.6\,\mu\text{m},0)$. 
(c) Corresponding time evolution of the longitudinal electric field $E_x$. 
(d) Time evolution of the magnetic field component $B_z$ at the same location.}
\label{figPoy} 
\end{figure}%
The physical mechanism underlying energy concentration is elucidated through Poynting vector analysis in Fig.~\ref{figPoy}. 
Spatial distribution of $P_y$ in Fig.~\ref{figPoy}(a) reveals pronounced energy convergence toward the central axis near the nanotube entrance. Temporal evolution at 100\,nm preceding the nanotube in Figs.~\ref{figPoy}(b) shows energy flow alternating between inward and outward directions with optical cycle periodicity, though inward components dominate significantly—producing net energy convergence. Magnetic field orientation is predominantly along $z$, while electric field exhibits both transverse and longitudinal components. As shown in Figs.~\ref{figPoy}(c) and \ref{figPoy}(d), the $E_x$ component, representing back-scattered radiation, displays a phase difference relative to $B_z$ (primarily from the incident field). This phase relationship creates the asymmetry between inward and outward energy flows. The phase shift between longitudinal and transverse field components is essential for beam convergence and divergence~\cite{Niziev2020JOSAA}. The saturation behavior observed in previous figures arises from non-resonant localized surface plasmon, similar to Mie scattering from nanoplasmas, becoming density-independent in the off-resonance regime. The complete process involves: incident laser field excitation of evanescent waves at the aperture rim, generating back-scattered longitudinal fields; interference between scattered and incident fields producing converging energy flow toward the laser axis; and coupling of concentrated energy into tubular plasma waveguide modes with moderate attenuation. 

The observed enhancement mechanism is governed primarily by interactions at the two open ends of the tubular structure, rather than by the lateral dimensions of the resulting overdense plasma. This insight suggests that our findings are generalizable to nanoholes in planar sheets. The nanotube geometry provides significant computational advantages by substantially reducing the required number of macro-particles while maintaining physical relevance. Since the transmission is sensitive to the ratio of inner radius to the wavelength, this plasma aperture can be exploited to design dichroic filters at relativistic intensities. 

\section{Conclusion}
In conclusion, we have demonstrated robust field enhancement of intense laser pulses in subwavelength plasma apertures using PIC simulations. The observed enhancement stems from localized surface plasmons, which efficiently couples incident radiation into guided modes within the plasma channel. This non-resonant effect saturates at plasma densities around $20n_c$ and is governed primarily by aperture geometry rather than material properties. Analysis reveals that energy concentration arises from convergent Poynting flow due to interference between incident pulses and scattered longitudinal fields. The mechanism persists at relativistic intensities, underscoring its robustness for high-field applications such as plasma dichroic filter.

\begin{acknowledgments}
This work was supported by Natural Science Foundation of Zhejiang Province (LY19A040005).
\end{acknowledgments}

\section*{Data Availability Statement}
The data that support the findings of this study are available from the corresponding author upon reasonable request.

%

\end{document}